# Grain boundary effect on nanoindentation: A multiscale discrete dislocation dynamics model

Songjiang Lu[a], Bo Zhang[a], Xiangyu Li[a], Junwen Zhao[b], Michael Zaiser[a,c]

Haidong Fan[d*], Xu Zhang[a*]

[a]*Applied Mechanics and Structure Safety Key Laboratory of Sichuan Province, School of Mechanics and Engineering, Southwest Jiaotong University, Chengdu 610031, China*

[b]*School of Materials Science and Engineering & Key Lab of Advanced Technologies of Materials (Ministry of Education), Southwest Jiaotong University, Chengdu 610031, China*

[c]*Department of Materials Science and Engineering, Institute of Materials Simulation WW8, FAU University of Erlangen-Nuremberg, Dr.-Mack-Str.77, 90762 Fürth, Germany*

[d]*Department of Mechanics, Sichuan University, Chengdu 610065, China*

*\*Correspondent authors E-mails: xzhang@swjtu.edu.cn (XZ), hfan85@scu.edu.cn (HF).*

**Abstract**

Nanoindentation is a convenient method to investigate the mechanical properties of materials on small scales by utilizing low loads and small indentation depths. However, the effect of grain boundaries (GB) on the nanoindentation response remains unclear and needs to be studied by investigating in detail the interactions between dislocations and GBs during nanoindentation. In the present work, we employ a three-dimensional multiscale modeling framework, which couples three-dimensional discrete dislocation dynamics (DDD) with the Finite Element method (FEM) to investigate GB effects on the nanoindentation behavior of an aluminum bicrystal. The interaction between dislocations and GB is physically modeled in terms of a penetrable GB, where piled-up dislocations can penetrate through the GB and dislocation debris at GBs can emit full dislocations into grains. In the simulation, we confirmed two experimentally observed phenomena, namely, pop-in events and the dependence of indentation hardness on the distance from GB. Two pop-in events were observed, of which the initial pop-in event is correlated with the activation and multiplication of dislocations, while the GB pop-in event results from dislocation transmission through the GB. By changing the distance between the indenter and GB, the simulation shows that the indentation hardness increases with decreasing GB-indenter distance. A quantitative model has been formulated which relates the dependency of indentation hardness on indentation depth and on GB-indenter distance to the back stress created by piled-up geometrically necessary dislocations in the plastic zone and to the additional constraint imposed by the GB on the plastic zone size.

## 1. Introduction

With the rapid development of nano-technology, the mechanical behavior of materials on micro and nano scales has become an important subject of scientific investigation, and differences between micro-scale and macroscopic mechanical properties have received increasing attention [1-3]. Indentation testing is one of the most convenient methods to investigate the mechanical response of materials on micro and nano scales because of its comparative simplicity, the possibility to probe material microstructures in small volumes near the surface of any bulk sample, and its capabilities in testing thin films where tensile experiments are difficult to perform [4].

In polycrystalline metals consisting of many grains that are separated by grain boundaries (GBs), plastic deformation is mainly mediated by dislocation glide. Therefore, dislocation microstructure evolution and the interaction between dislocations and GBs plays a significant role in controlling the strength and deformation properties of metal polycrystals. The GB-dislocation interaction mechanisms are most conveniently investigated by studying bicrystals, where the existence of only one GB simplifies the problem to a significant extent [5-11]. Soer et al. [6,7] conducted nanoindentation experiments on Fe-Si bicrystals and Mo bicrystals, and observed that the load-depth curves were punctuated by two displacement jumps ("pop-in events") which they interpreted as respective signatures of grain interior yielding and of dislocation transmission through the GB. Theoretically, these authors used a gradient plasticity framework in conjunction with a Hall-Petch approach to analyze the deformation curves and to establish criteria for slip transmission across GBs [7]. Zhang et al. [8] carried out nanoindentation on copper bicrystals and found that the hardness first increases and then decreases with increasing indentation depth. They pointed out that this phenomenon is related to the interaction between dislocations and GB. Due to the resistance of the GB to dislocation motion, dislocations accumulating in front of the GB cause a back stress in the grain interior, so that a kinematic hardening effect can be observed. However, as the indentation depth increases, the accumulated dislocations penetrate through the GB once the resolved shear stress acting on the dislocations exceeds a critical value and then expand freely in the adjacent grain, leading to softening. From the aforementioned experimental results, it can be clearly seen that the GB plays a significant role in the mechanical response of bicrystals, and by extension of polycrystalline materials.

In nanoindentation experiments, various techniques have been used to characterize the dislocation configuration around the indented area, such as Transmission Electron Microscopy [12, 13] and newly-developed Electron Channeling Contrast Imaging under controlled diffraction conditions [14]. However, it is still difficult to quantitatively analyze the dislocation distribution or to monitor its evolution during deformation *in situ*. Therefore, numerical simulation remains an excellent tool to investigate those aspects of plasticity during nanoindentation that cannot be directly accessed by experiment, and to interpret easily accessible experimental information (such as load displacement curves) in terms of microstructural mechanisms. Currently, the crystal plasticity finite element method (CPFEM) and molecular dynamics are the main tools used for nanoindentation simulations. Li et al. [15] performed CPFEM simulations to investigate the effects of grain orientation and grain geometry on the indentation response of a polycrystalline 2024 aluminum alloy. The simulation results showed that stress and misorientation distributions are continuous across low angle GBs, while high angle GBs act as strong barriers for plastic slip and lead to stress concentration at the GBs. Liu et al. [9] reported that the indentation responses of copper single crystals and copper bicrystals in CPFEM simulations were almost identical. This illustrates an important limitation of the above-mentioned crystal plasticity models, which only considered the different crystallographic orientations in different grains but not the different processes which govern plasticity in the bulk and near GBs on the microstructure level. Thus, the most important effects of GBs on the indentation response of materials could not be captured, and therefore, a different modelling approach, which adequately accounts for the relevant physical processes at and near GBs, is required. Compared with CPFEM simulation, DDD simulation has great advantages in this respect since it provides complete information regarding the evolution of dislocation microstructures and thus can account for differences between dislocation behavior in the bulk and near GBs. Because DDD simulations of nanoindentation need to deal with inherently nonlinear contact problems associated with the evolving substrate-indenter contact, simulation schemes must be used that are capable of dealing with general boundary conditions. Indentation size effects and influences of indenter shape were investigated by two-dimensional DDD simulations on single crystals [16, 17], by using the method introduced by Needleman and Van der Giessen [18] which couples DDD with FEM to evaluate dislocation interactions in terms of surface-corrected infinite-body dislocation stress fields. Two-dimensional DDD simulations on polycrystals also captured the effects of grain size and indentation depth on indentation response [19, 20]. The first three-dimensional (3D) DDD simulation of nanoindentation was conducted by Fivel et al. [12]. They used a DDD model with coarse-grained dislocation cores, which they coupled with FEM in order to handle the contact boundary value problem and study the plastic zone of indentation. Recently, Hu et al. [21] used similar methodology to investigate the correlation of incipient plasticity with dislocation nucleation and multiplication during indentation. Po et al. [22] used a scheme that couples DDD with the boundary element method to simulate nanoindentation of copper single crystals, and studied the effect of indenter shape on indentation hardness and dislocation structure. However, to the best knowledge of the authors, few studies have so far investigated the effect of GB on nanoindentation by using 3D DDD, which requires the introduction of an adequate GB model into the 3D DDD framework. Therefore, our focus is on the simulation of bicrystal nanoindentation, using a framework which couples FEM and 3D DDD with a penetrable GB model. The simulation aims at revealing the characteristics of dislocation evolution and the mechanisms of GB-dislocation interaction during nanoindentation, in order to understand better the effects of GB-dislocation interaction on the mechanical response of polycrystalline materials in general.

This paper is organized as follows. First, the multiscale discrete dislocation plasticity model and GB model are briefly described. Second, indentation simulations on bicrystal events are performed to analyze the initial pop-in event, GB pop-in event and the variation of hardness versus distance to GB. Then, we use the simulation data to construct a phenomenological model using a Hall-Petch type relation in conjunction with the spatio-temporal evolution of geometrically necessary dislocation density to analyze the GB hardening effect in terms of a pile-up model. In conclusion, the main findings are summarized.

## 2. Multiscale dislocation dynamics framework and GB model

DDD allows to analyze plastic deformation mechanisms during nanoindentation by tracking the evolution of the dislocation microstructure. However, DDD was originally developed as a bulk simulation method and encounters challenges when dealing with complex boundary conditions. This poses particular problems in nanoindentation simulations because of the inherently nonlinear contact problem. On the other hand, FEM is a convenient simulation method to solve complicated boundary conditions. Therefore, a hybrid (multiscale) model that couples DDD with FEM is the method of choice [12, 23, 24]. In this work, the 3D discrete-continuous model (DCM) initially developed by Zbib et al. [25, 26] and improved by Jamond et. al. [27] and Huang et al. [28] is used. This model has several advantages [28]. The elastic-plastic problem is treated in a multiscale framework where the (discontinuous) plastic strain obtained from the DDD simulation is coarse grained to yield smoothly varying stress and plastic strain fields on the FEM scale. At the same time, in the DDD simulation, the thus evaluated coarse-grained FEM stress fields are corrected to account for the high stress fields close to the dislocation lines and thus to correctly represent the interaction of nearby dislocations, which is underestimated in traditional DCM approaches. [25,26,29-31]. In this work, use the DCM famework described in Ref. [29]

which we generalize, in order to investigate GB effects in the indentation responses of bicrystals, by introducing a 3D penetrable GB model into the DDD calculations.

*2.1. Basic idea of multiscale coupling in DCM*

The basic idea of multiscale DCM as introduced in [27,28] can be summarized as follows: In 3D DDD, the stress field exerted by the dislocation ensemble on a segment *n* is normally represented as a sum over segment stress fields,

$$\sigma(r_n) = \sum_{i \neq n} \sigma^s(b_i, t_i, r_n - r_i), \tag{1}$$

where the segment stress fields $\sigma^s(b_i, t_i, r - r_i)$ depend on the segment Burgers vectors $b_i$ and tangent vectors $t_i$. The segment stress fields are highly singular: They decay like $1/r^2$. This implies that they exhibit two singularities, one at zero – which renders the time integration of the interactions of nearby dislocations computationally very expensive – and one on the infinite boundary – which makes the summation over the stresses of distant segments computationally expensive since it is not *a priori* possible to truncate.

Multiscale DCM resolves this problem by introducing a coarse grained stress field $\sigma_a(r)$. This represents the stress field of a coarse grained dislocation system where, by means of a coarse graining function $w_a(r')$, the Burgers vector of each dislocation segment has been distributed isotropically over a region with radius of the order of *a* around the centerpoint of the segment. (For generalization to anisotropic coarse graining schemes see [28].) The coarse grained stress field correctly captures the structure of the dislocation stress field on scales much larger than *a*, but due to the coarse graining it misses the high stresses near the dislocation line singularities and therefore cannot correctly describe the interactions of nearby dislocations. The true stress field can, however, be represented without approximation as the sum of the coarse grained stress field $\sigma_a(r)$ which describes long-range interactions, and the residual stress field $\sigma_{SR}(r) := \sigma(r) - \sigma_a(r)$ which accounts for short-range interactions. The point of this decomposition is that the contribution $\sigma_a(r)$ can be evaluated in two different manners:

(i) In evaluating $\sigma_{SR}(r)$, one uses a particular functional form for the coarse graining function proposed by Cai et. al. [32] that allows to compute the stress $\sigma_a(r_n)$ on a segment $n$ as a sum of coarse grained segment stress fields $\sigma_a^s(b_i, t_i, a, r_n - r_i)$ that can be evaluated analytically (for explicit expressions see Ref. [32]), such that one can write

$$\sigma_{SR}(r_k) = \sum_{i \neq n} \sigma_{SR}^s(b_i, t_i, r_n - r_i) = \sum_{i \neq n} \left[ \sigma^s(b_i, t_i, r_n - r_i) - \sigma_a^s(b_i, t_i, a, r_n - r_i) \right] \quad . \tag{2}$$

An important benefit of this procedure is that, as shown by Cai et. al. [32], the difference $\sigma_{SR}^s(b_i, t_i, r_n - r_i)$ of the singular and the coarse- grained stress field of a segment is a short-ranged function that decreases rapidly for $|r_n - r_i| \gg a$. Therefore one may restrict, in Eq. (2), the summation to a finite domain. In the present work we take this domain to be a sphere $S_n$ of radius $2a$ around $r_n$. The situation is depicted schematically in Fig.1, right.

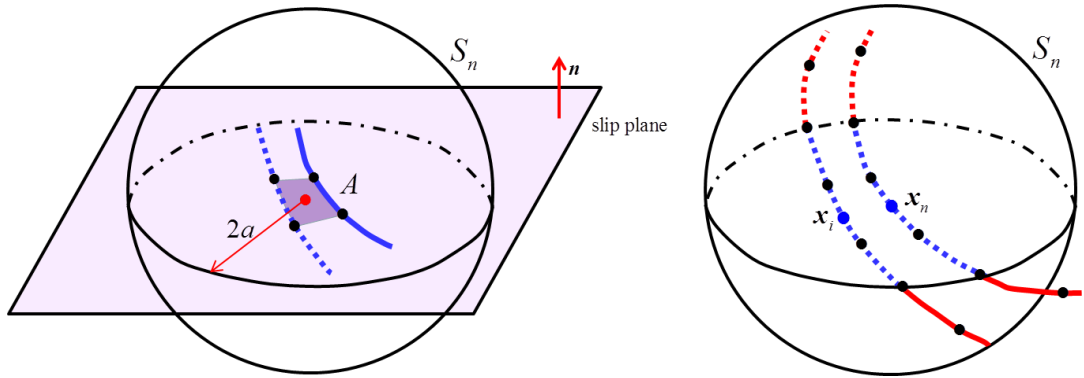

**Fig. 1.** Schematic illustration: (left) sphere $S_n$ in which the incremental plastic strain caused by slip of segment $n$ is coarse grained according to $w_a(r)$ for use as eigenstrain in the FEM stress calculation; (right) calculation of the stress at the middle point of dislocation segment $n$ by adding the short-ranged stresses of segments within $S_n$, (blue segments) to the FEM stress, after [28].

(ii) For evaluating the long range stress interactions which are described by the field $\sigma_a(r)$ alone, a different procedure is required. Direct summation over segment stress fields would run into the difficulty that $\sigma_a(r)$ decays slowly like $1/r^2$ (hence, the summation would need to be done over all segments in the simulation) and is furthermore influenced by boundary conditions, such that for general boundary conditions an analytical expression for the long-ranged part is not available. Instead, one uses the fact that the calculation of any dislocation stress field can also be envisaged as the solution of an eigenstrain problem. The differential Eigenstrain created by a dislocation segment $n$ of Burgers vector $\boldsymbol{b}_n$ sweeping an area d$A$ on its slip plane with normal $\boldsymbol{n}_n$ is given by

$$d\varepsilon^{\mathrm{p}}(r) = \frac{dA}{2}(\boldsymbol{n}_n \otimes \boldsymbol{b}_n + \boldsymbol{b}_n \otimes \boldsymbol{n}_n)\delta(r - r_n), \qquad (3)$$

the overall incremental eigenstrain is obtained by summation over all segments, and the overall eigenstrain is obtained by summation of the incremental eigenstrains over all time steps. We now define a coarse grained eigenstrain where we spread the Burgers vector of the dislocation, using the same coarse graining function $w_a(r)$ as used for direct evaluation of $\sigma_a(r)$. Furthermore, we restrict the evaluation of the coarse grained eigenstrain to the sphere $S_n$, which ensures both numerical efficiency and conceptual consistency of the coarse graining procedure. Thus we evaluate the coarse grained eigenstrain created by a segment sweeping the area d$A$ as

$$d\varepsilon_a^p(r) = N\frac{dA}{2}(\boldsymbol{n}_n \otimes \boldsymbol{b}_n + \boldsymbol{b}_n \otimes \boldsymbol{n}_n)w(r - r_n)\Theta_{S_n}(r - r_n) \qquad (4)$$

where the function $\Theta_{S_n}(\boldsymbol{r})$ is one if r is contained within $S_n$ and zero otherwise (Fig 1, left). The normalization factor $N$ where $N^{-1} = \int_{S_n} w(\boldsymbol{r})d^3r$ ensures that the volume integrals of the coarse grained and the original eigenstrain coincide [28]. Again, the overall coarse grained eigenstrain is obtained by summation of the differential eigenstrains over all segments and time steps.

Finally, the thus evaluated eigenstrain is used to determine, using standard Finite Element methodology, the coarse grained stress field $\sigma_a(\boldsymbol{r})$ in a manner that accounts naturally for the imposed boundary tractions and/or displacements.

*2.2. DDD-FEM coupling: technical aspects*

The present DCM model contains two modules, i.e. the DDD and FEM. The DDD module handles the dynamical evolution of dislocations and calculates the resulting plastic strain, which is then coarse grained and distributed on the Gauss integration points (GIPs) of the FEM mesh. In the FEM module, we treat the plastic strain $\varepsilon_p^a$ transferred from the DDD module and the stress $\sigma_0$ induced by initial dislocations as the eigenstrain and pre-stress field, respectively. The stress state of GIPs is described through the constitutive model,

$$\sigma_a^{\text{GIP}} = \boldsymbol{C}(\boldsymbol{\varepsilon} - \boldsymbol{\varepsilon}_p^a) + \boldsymbol{\sigma}_0,$$

where $\boldsymbol{C}$ is the elastic stiffness matrix. The displacement, strain and stress of any given point in the sample are then determined by solving the boundary-value problem in commercial FEM software Ansys®. Finally, the thus evaluated stress field is interpolated back to the segment centerpoints as

$$\sigma_n^a(\boldsymbol{r}_n) = \sum \sigma_i^{\text{node}} N_i(\boldsymbol{r}_n),$$

where $\sigma_i^{\text{node}}$ are the FEM stresses at the nodes located in the same element as the evaluated dislocation segment midpoint $\boldsymbol{r}_n$, $N_i$ is the interpolation function of the adopted finite element. For a hexahedral element as used in our nanoindentation simulation, $N_i$ is defined as

$$N_i = \frac{1}{8}(1+\xi_i\xi)(1+\eta_i\eta)(1+\zeta_i\zeta) \qquad (i = 1, 2, \cdots, 8), \tag{1}$$

where $\xi_i, \eta_i, \zeta_i$ are the local coordinate of the nodes in the element, and $\xi, \eta, \zeta$ are the local coordinates of the dislocation segment midpoint, as shown in Fig. 2. The local coordinate ($\xi, \eta, \zeta$) is related to the global coordinate ($x, y, z$) as

$$\xi = 2(x-x_O)/l_x, \quad \eta = 2(y-y_O)/l_y, \quad \zeta = 2(z-z_O)/l_z, \tag{2}$$

where $x_O, y_O, z_O$ are the global coordinate of the element center, and $l_x, l_y, l_z$ are the element sizes in three dirctions, see Fig. 2.

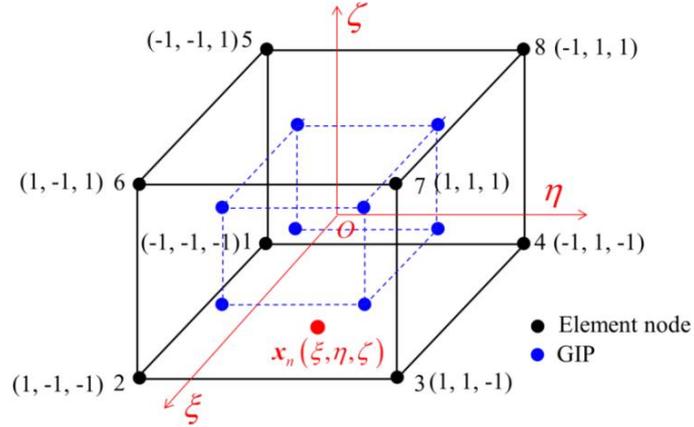

**Fig. 2.** A hexahedral element in local coordinate system ($\xi, \eta, \zeta$).

Finally, the total stress which drives dislocation motion is obtained by adding the short-range stress

$$\boldsymbol{\sigma}_n = \boldsymbol{\sigma}_n^{FE}(\rho_n) + \sum_{i \neq n \in S_n} \boldsymbol{\sigma}_i^{SR}(\boldsymbol{r}_n - \boldsymbol{r}_i), \tag{3}$$

Finally, we make a line-tension correction $\boldsymbol{F}_n^{LT}$ to correctly represent the interaction of adjacent segmetns of the same line, and thus we obtain the total force on the dislocation as

$$\boldsymbol{F}_n^{tot} = \boldsymbol{F}_n^{PK} + \boldsymbol{F}_n^{LT} = (\boldsymbol{\sigma}(\boldsymbol{r}_n) \cdot \boldsymbol{b}_n) \times \boldsymbol{\xi}_n + \boldsymbol{F}_n^{LT}, \tag{4}$$

The velocity of the dislocation is determined from this force assuming a linear mobility law:

$$v = \frac{\boldsymbol{F}_{\text{tot}} \cdot \boldsymbol{t}}{B}, \tag{5}$$

where $\boldsymbol{t} = \boldsymbol{n} \times \boldsymbol{\xi}$ is the dislocation glide direction and $B$ is the viscous drag coefficient, which is taken to be $10^{-4}$ Pa·s in this work [28].

*2.3 Computational efficiency of the DCM model*

DCM as described here is, from the point of view of computational efficiency, superior to standard 3D DDD when it comes to simulating large systems with general boundary conditions. The use of FEM to evaluate the long-range dislocation stress fields makes it not only possible to handle general boundary conditions. It also allows to truncate the summation over segment stress fields, thus reducing the computational cost of evaluating segment-segment interactions to order *N* while still being able to use simple analytical expressions in the summation. At the same time, the fact that we use coarse grained eigenstrains in the FEM calculations results in a slow spatio-temporal variation of the FEM stress as compared to the interaction stresses of nearby dislocations. Therefore, the FE time step can be increased to up to 40 times the DDD time step (one FE step per 40 DDD steps) without compromising the accuracy of the simulations. The length- and time scale separation between DDD and FEM thus significantly enhances the computational efficiency of DCM: On the one hand, the number time-consuming FE calculations can be restricted, on the other hand, in DDD the time-consuming summation over the interaction stresses of distant dislocation segments is avoided since these interactions are accounted for by the FEM stress field.

*2.4. GB model*

In the present work, we adopt a 3D penetrable tilt GB model proposed by Fan et al. [5] to handle the interactions between dislocations and GBs. In this model, two GB-dislocation interaction scenarios were considered, namely, dislocation penetration through GBs and dislocation emission from GBs. As shown in Fig. 3, two adjacent grains with misorientation angle $\theta$ are separated by a symmetric tilt boundary, such that the slip planes I and II have a common line of intersection with the GB.

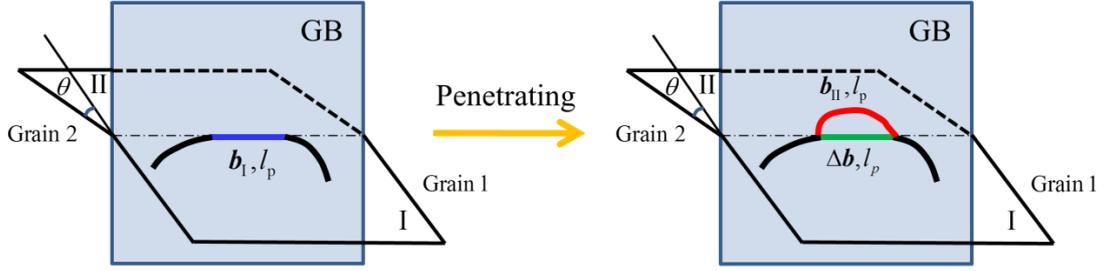

**Fig. 3.** Schematic diagram showing a dislocation penetrating through a grain boundary (GB).

When approaching the GB, an incoming dislocation segment (colored in blue) with Burgers vector $\boldsymbol{b}_\text{I}$ and length $l_p$ on slip plane I in Grain 1 is initially blocked in front of the GB. However, if the stress acting on the incoming dislocation segment $\boldsymbol{b}_\text{I}$ exceeds a critical stress $\tau_\text{CRSS}$, the segment is transmitted through the GB from Grain 1 to Grain 2 and becomes a new dislocation segment (marked in red) with Burgers vector $\boldsymbol{b}_\text{II}$ and length $l_p$ on slip plane II of Grain 2. In addition, considering the conservation of Burgers vector, a debris dislocation $\Delta\boldsymbol{b} = \boldsymbol{b}_\text{I} - \boldsymbol{b}_\text{II}$ (colored in green) with the same length $l_p$ is left on the GB after the transmission event. Fig. 4 shows a two-dimensional schematic of the dislocation transmission process in a plane perpendicular to the tilt axis of the GB. The direction unit vector of the tilt axis is denoted as $t$, and the dislocation Burgers vector can be decomposed into a vector parallel and a vector perpendicular to the tilt axis, $\boldsymbol{b} = \boldsymbol{b}^\perp + \boldsymbol{b}^\parallel$, where the parallel vector has the length $|\boldsymbol{b}^\parallel| = b^\parallel = \boldsymbol{b}\cdot\boldsymbol{t}$ and the perpendicular vector $\boldsymbol{b}^\perp = \boldsymbol{b} - t b^\parallel$ has length $b^\perp$. The magnitude of the dislocation debris Burgers vector can then be easily determined from the figure as

$$\Delta b = 2 b_{\mathrm{II}}^\perp \sin\left(\frac{\theta}{2}\right) \qquad (6)$$

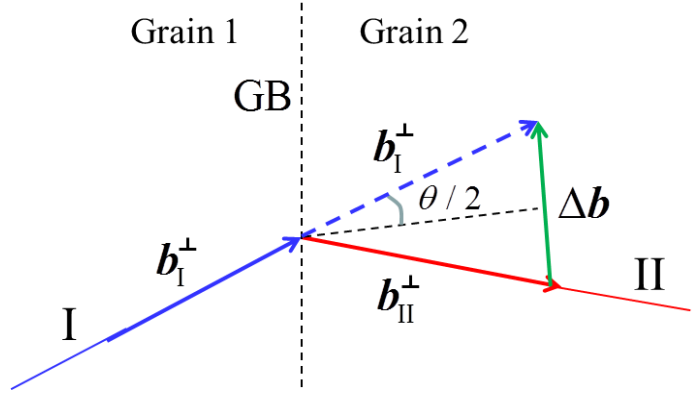

**Fig. 4.** Two-dimensional schematic of dislocation transmission.

The critical resolved shear stress $\tau_{\mathrm{CRSS}}$ can be calculated from the energy conservation law as

$$\tau_{\mathrm{CRSS}} b_{\mathrm{I}}^2 = E_{\mathrm{gb}} b_{\mathrm{I}} + \alpha G (\Delta b)^2, \qquad (7)$$

where the first term on the right-hand side represents the energy increase of the grain boundary with $E_{\mathrm{gb}}$ being the GB energy density. The term $\alpha G(\Delta b)^2$ represents the elastic energy associated with the GB ledge caused by the dislocation debris left on the grain boundary, with $\alpha$ being a non-dimensional constant accounting for the atomistic configuration of the dislocation debris and $G$ the shear modulus. According to Hasson et al.'s measurements [31], the GB energy density $E_{\mathrm{gb}}$ in Eq. (7) is almost a constant when $\theta \in (\theta_1, \theta_2]$, while for $\theta$ close to $0°$ and $90°$, $E_{gb}$ changes linearly with $\theta$. Therefore, $E_{gb}$ can be approximately expressed by the following equation [33]:

$$E_{gb} = \begin{cases} k\theta/\theta_1 & 0° < \theta \leqslant \theta_1 \\ k & \theta_1 < \theta \leqslant \theta_2 \\ k(90-\theta)/(90-\theta_2) & \theta_2 < \theta \leqslant 90° \end{cases}. \qquad (8)$$

where the parameters $k$, $\theta_1$, $\theta_2$ are 600 mJ/m², 20°, 70°, respectively, as used in Ref. [33]. In the present paper, the misorientation angle $\theta$ is set as $5°$. Therefore, $E_{gb}$ can be expressed as

$$E_{gb} = k\theta/\theta_1. \qquad (9)$$

Combining Eqs. (6),(7), (9) and considering that $b_I = b_{II} = b$, the critical resolved shear stress for dislocation penetration can be expressed as

$$\tau_{CRSS} = \frac{k\theta}{\theta_1 b} + 4\alpha^* G \sin^2(\frac{\theta}{2}). \qquad (10)$$

where $(\alpha^*)^2 = \alpha^2(1-(\boldsymbol{b}.\boldsymbol{t})^2)$. With increasing number of dislocation penetration events, the energy of the dislocation debris on the GB also increases and a mechanism to release the corresponding energy is needed. It has been experimentally confirmed that steps and ledges on GBs are important nucleation sites for dislocations [34]. Such a nucleation/emission event is schematically illustrated in Fig. 5: the dislocation debris (colored in green) with total Burgers vector

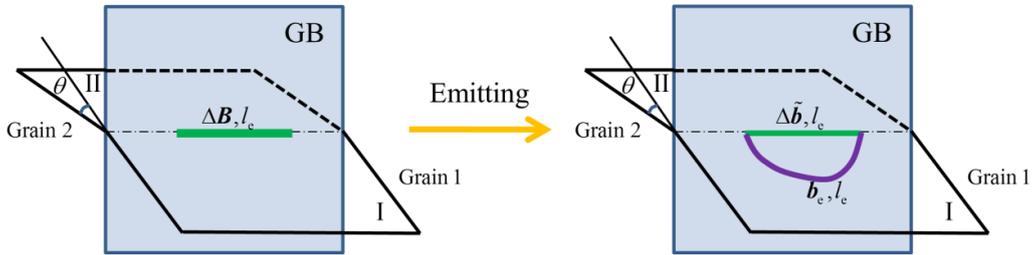

**Fig. 5.** Schematic of dislocation debris decomposition.

$\Delta \boldsymbol{B} = \sum \Delta \boldsymbol{b}$ and length $l_e$

emits a perfect lattice dislocation $b_e$ (colored in purple) with the same length $l_e$ in the slip plane of one of the grains. After the emission event, the Burgers vector of new dislocation debris on GB becomes $\Delta \tilde{b}$ with length $l_e$. Similar to the penetration event, the conservation of Burgers vector must be satisfied, i.e., $\Delta B = \Delta \tilde{b} + b_e$. Furthermore, the dislocation emission process must be energetically favorable,

$$\Delta B^2 \geq \Delta \tilde{b}^2 + b_e^2. \tag{11}$$

Dislocation emission occurs as soon as any Burgers vector $b_e$ consistent with Burgers vector conservation fulfills this inequality. Based on these rules, the grain and the slip plane of the emitted dislocation $b_e$ can be determined.

## 3. Simulation setup

The FEM indentation model (generated by Ansys®) is shown in Fig. 6(a). We simulate indentation with a Berkovich-type diamond indenter with a half-angle of $65.3°$. Since the elastic modulus of the indenter is much higher than that of the indented material, the indenter is in the simulation modeled as a rigid body. One side of the indenter is assumed to run parallel to the intersection line of GB plane and indented surface, as shown in Fig. 6(a). The dimensions of the indented sample in *X*, *Y* and *Z* directions are $l_x$=1400*b*, $l_y$=1200*b* and $l_z$=600*b*, respectively, where *b* is the magnitude of Burgers vector and chosen as 0.25 nm in this paper. Furthermore, the height of the indented sample is set to be ten times larger than the maximum indentation depth of 44*b*, such as to make the indentation response approximately independent of model size [36]. 20-node hexahedral elements (SOLID 186 in ANSYS) were used to improve the accuracy of the FE calculation. As the nanoindentation simulation involves a contact problem and the contact area between indenter and indented material is relatively small, the FE mesh underneath the indented region was refined (see Fig. 6(a)). All FE nodes on the bottom surface of the sample are completely constrained, while the four surrounding surfaces are free. The displacement of the indenter along the *z* direction is increased at fixed rate. Contact between indenter and sample is assumed rigid.

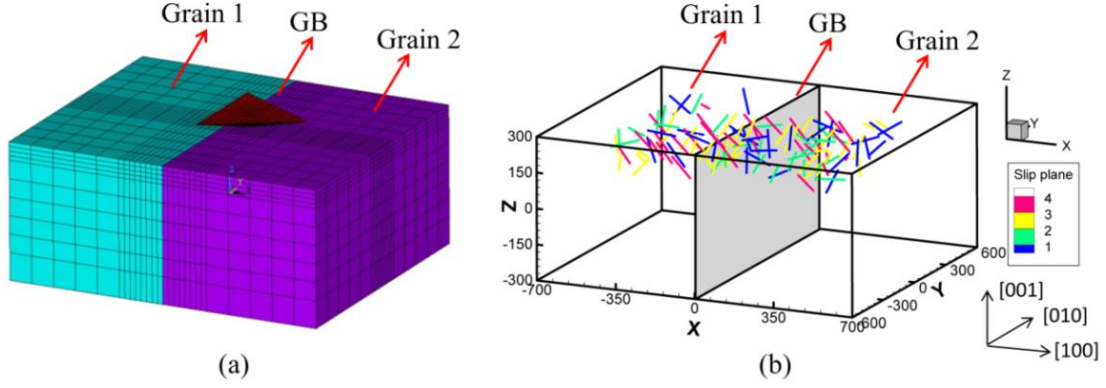

**Fig. 6.** Indentation model. (a) finite element model (generated by commercial software Ansys), (b) initial configuration of the discrete dislocation dynamics model (generated by software Tecplot).

The GB plane is vertically located in the middle of the model (see Fig. 6(b)), the misorientation between the two grains of the bicrystal is set to $\theta = 5°$. The indentation direction is aligned with the [001] direction of the crystal lattice of the indented grain as shown in Fig. 6(b). To model dislocation nucleation underneath the indenter, we create 120 dislocation segments with pinned endpoints (10 segments per slip system), and place them at random locations within a thin layer underneath the specimen surface with $X \in [-500b, 500b]$, $Y \in [-400b, 400b]$ and $Z \in [100b, 300b]$ as shown in Fig. 6(b). The segment orientations are random with uniform distribution within the respective slip planes. The segment lengths, which determine the critical stresses for dislocation nucleation, are chosen in such a manner that the hardness values obtained from our simulations match those obtained in large scale MD simulations of indentation in absence of bulk dislocation sources [37]. We use a uniform distribution of lengths between $80b$ and $100b$, which results for an indentation depth of $40b$ in typical hardness values $0.2\mu < H < 0.25\mu$, in agreement with our reference simulations [37].

The indentation simulation considers displacement-controlled loading with the indenter displacement rate $\dot{h}$ chosen such as to impose a constant indentation strain rate $\dot{h}/h = 1 \times 10^4$ s$^{-1}$. In the simulations, typical material parameters of aluminum bicrystals are used. These parameters are listed in Table 1.

**Table 1.** Parameters of aluminum bicrystal used in simulations.

| Material parameters | Symbol | Value |
|---|---|---|
| Density (kg/m$^3$) | $\rho$ | 2700 |
| Shear modulus (GPa) | $G$ | 26 |
| Poisson's ratio | $v$ | 0.345 |
| Burgers vector magnitude (nm) | $b$ | 0.25 |
| Viscous drag coefficient (Pa·s) | $B$ | $10^{-4}$ |
| Grain boundary (GB) parameter | $\alpha$ | 0.5 |
| GB parameter (mJ/m$^2$) | $k$ | 600 |
| GB parameter (°) | $\theta_1$ | 20 |
| GB parameter (°) | $\theta_2$ | 70 |

## 4. Results

### 4.1. Pop-in event

During nanoindentation, discontinuities on the load-depth curves can be observed, which are related to the first activation and multiplication of dislocations under the indenter tip. These events indicate the transition from elastic to elasto-plastic response of the indented material. In the load-controlled loading mode, the load-depth curves display a displacement excursion known as initial "pop-in" phenomenon, whereas an abrupt load drop is observed in the case of displacement-controlled loading [38, 39]. In order to avoid artefacts due to the stochastic nature of the plastic response in dislocation dynamics simulations and to ensure that observed pop-in like features constitute generic characteristics of indentation behavior rather than mere fluctuations, three simulations with a distance from indenter to GB of 50$b$ were conducted. These three simulations use different initial dislocation configurations but the same initial dislocation density and average dislocation length. Pop-in events were observed in all three cases, as shown in Fig. 7, where it can be seen that the pop-in events on the load-depth curves (Figs. 7g, 7h and 7i) all appear at the first dislocation activation (Figs. 7d, 7e and 7f). This indicates that the initial pop-in is associated with the activation of dislocation sources and the subsequent dislocation multiplication [13, 40], as observed in previous experiments [40,41]. It should be pointed out that although the indentation loading was realized in displacement-controlled mode, the load-depth curves show a pop-in phenomenon rather than a sudden load drop. The absence of a load drop probably results from the low data output frequency. In the simulation framework, the simulation results are exported every 40 steps, thus some significant data points may be neglected.

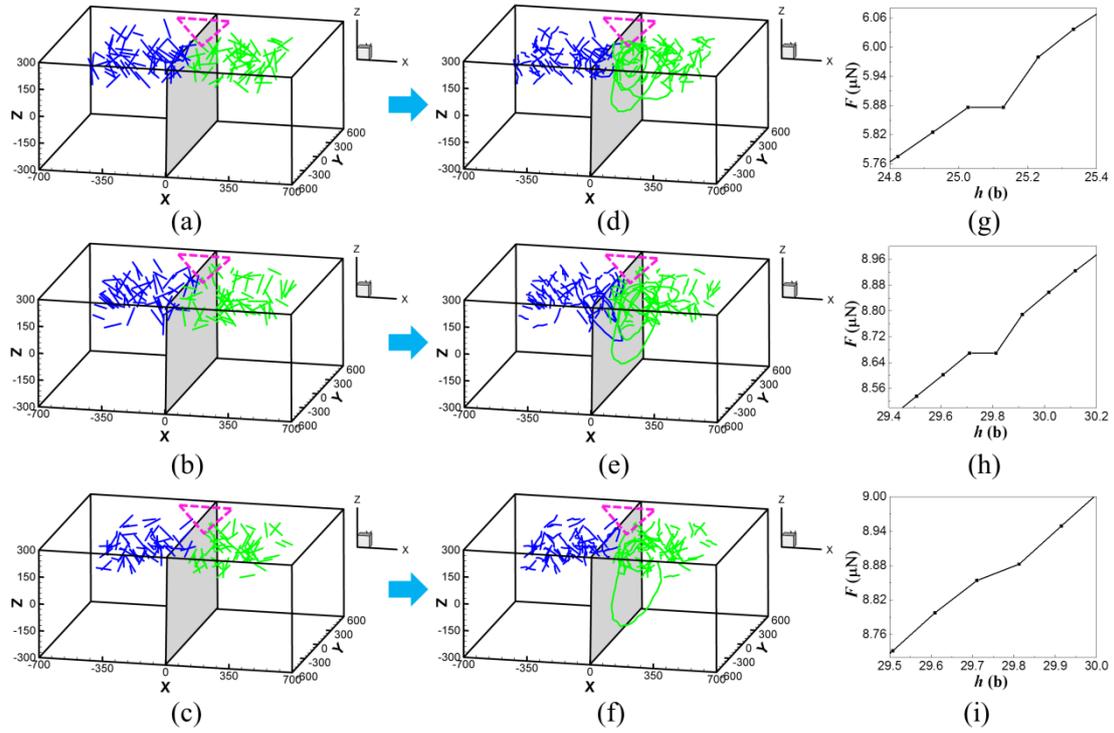

**Fig. 7.** Dislocation structure before and after initial pop-in event. (a–c) Initial dislocation arrangements, (d–f) dislocation arrangements after activation of the first dislocation source and (g–i) enlarged load-depth curves at the initial pop-in event for the three initial dislocation structures. The indenter-GB distance is 50$b$. The position of the indenter is schematically marked with a pink dotted triangle.

Besides the initial pop-in event resulting from the first activation of dislocations, a displacement burst known as GB pop-in can also be observed when dislocations penetrate through GBs. Fig. 8 shows the dislocation structure before and after such a penetration event, and the corresponding load-depth curve, again with a GB-indenter distance of 50$b$. As shown in Fig. 8(a), a large number of dislocations accumulate in front of the GB, which suppresses the further movement of dislocations during loading. However, as the indentation depth and the stress acting on the GB gradually increase, the dislocations are transmitted through the GB and leave dislocation debris (red lines in Fig. 8(b)) on the GB once the resolved shear stress acting on a trapped dislocation reaches the critical value for dislocation transmission, Eq. (15).

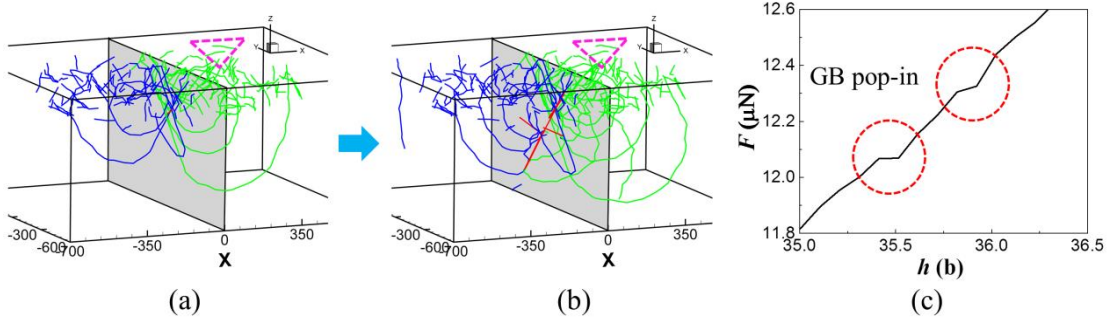

**Fig. 8.** Process of dislocation penetration through GB. The position of the indenter tip is schematically marked with a pink dotted triangle. (a) Dislocation structure before penetration, (b) dislocation structure after penetration, (c) load-depth curve around the moment of penetration.

As shown in Fig. 8(c), at the moment of dislocation transmission through the GB, a GB pop-in phenomenon appears on the load-depth curve due to the sudden release of stress underneath the indenter. In nanoindentation experiments, similar pop-in events on the load-depth curves have been observed when a sufficient number of dislocations simultaneously penetrate through a GB [11, 42], which qualitatively agrees with our simulation results. Similar to the initial pop-in event, the GB pop-in events obtained from simulations are not as pronounced as in experiment, since the number of dislocations simultaneously transmitted and the dimension of the simulated model are much smaller than in typical experiments.

Besides the pop-in phenomena coming from the transition from elastic deformation to elasto-plastic deformation and from dislocation transmission through GB, the intermittent operation and multiplication of dislocation sources can also induce subsequent pop-in events or fluctuation phenomena on the load-depth curve, as shown in Fig. 9.

In summary, from the above analysis it can be inferred that the essence of pop-in phenomena is the motion of a large number of dislocations, which generally occurs at the moment of initial yield (first dislocation source activation) of the material and at the moment when dislocations are first transmitted through GB and expand into the adjacent grain.

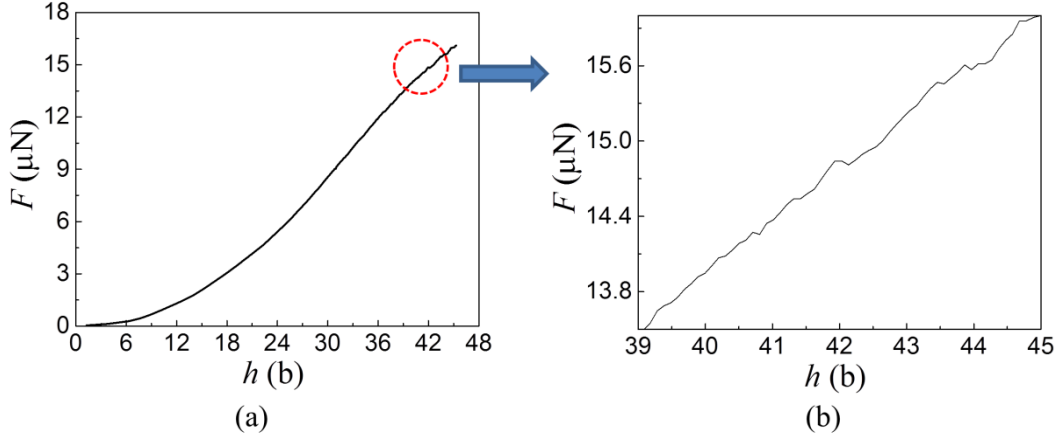

**Fig. 9.** Fluctuation phenomena on a load-depth curve as a result of intermittent operation and multiplication of dislocation sources. (a) Whole load-depth curve, (b) Enlarged part (red dashed circle in (a)) of the load depth curve.

*4.2. GB influence on indentation hardness*

In order to obtain reliable results, five simulations with different initial dislocation structures (initial pinned line segments are randomly distributed underneath the surface, with the segment length statistics and overall dislocation length being the same) were performed and the resulting load-depth curves were then averaged to get the final hardness value. In order to separate the dependence of indentation hardness on indentation depth from the dependence of indentation hardness on the distance to GB, two typical indentation depths of $32b$ and $42b$ were chosen to study the GB effect on hardness. Fig. 10 shows the relationship between indentation hardness and the distance from indenter to GB at the indentation depths of $32b$ and $42b$. In this figure, seven different distances $d$ have been evaluated, i.e., $d=-150b$, $d=-100b$, $d=-50b$, $d=0$ (on the GB), $d=50b$, $d=100b$ and $d=150b$. The hardness of the indented material is defined as $H = P/A$, where $P$ is the indentation load and $A$ is the contact area between indenter and bicrystal. It can be clearly seen from Fig. 10 that the hardness increases with decreasing GB-indenter distance, showing a significant GB hardening effect. Zhang et al. [9] and Soer et al. [7] have performed nanoindentation experiments on a copper bicrystal and a niobium bicrystal. Both of their experimental results show the trend of increasing hardness with decreasing GB-indenter distance, which is in good qualitative agreement with our simulated results.

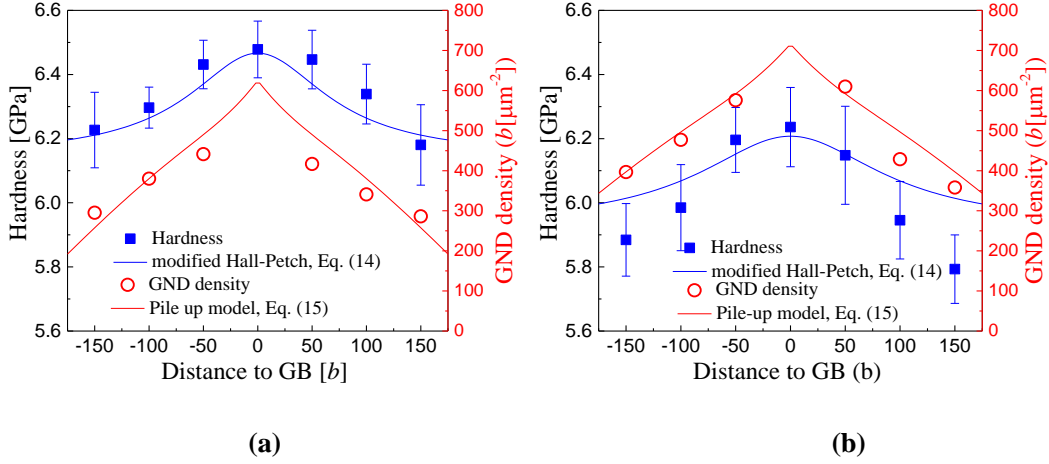

**Fig. 10.** Blue squares: Hardness vs. distance $d$ to GB, simulation data; blue line: modified Hall-Petch law, Eqs. (13) and (14); red circles: GND density $\rho_{\text{GND}}(V_d) = (1/2)\langle |\alpha_{32}| + |\alpha_{23}| \rangle_{V_d}$, averaged over the volume $V_d$ comprised between indenter tip and GB, vs. distance $d$ to GB, red line: circular pile-up model, Eqs. (15) and (A12); (a) and (b) represent results at indentation depths of $32b$ and $42b$, respectively; error bars represent standard deviation of five realizations.

## 5. Discussion

In order to provide a quantitative interpretation of the GB hardening phenomenon observed in our simulations (Fig. 10), we determine the spatial distribution of geometrically necessary dislocations (GND) underneath the indenter. We use this information to establish a mathematical model that allows us not only to evaluate the dependence of hardness on distance to GB and on indentation depth, but also evaluate the GND density between GB and indenter in quantitative terms.

*5.1. GND distribution*

One of the most significant advantages of DDD simulation is that it provides *in-situ* information about the evolution of the dislocation structure. From the discrete dislocation arrangement, a coarse-grained GND density tensor can be obtained by using the method proposed by Aifantis at al. [43]. This method has been adopted by Zhang et al. [43] to investigate how the internal length scale in strain gradient theory relates to parameters of the dislocation microstructure. The GND tensor along the *x* axis (averaged over the sample *y* and *z* directions) can be determined from the following equation [43]:

$$\boldsymbol{\alpha}(x) = \left\langle \frac{1}{\Delta V(x)} \sum_{i \in \Delta V(x)} \boldsymbol{I}_i \otimes \boldsymbol{b}_i \right\rangle, \tag{12}$$

where $\Delta V(x)$ is the volume of a slice that is centered at *x* and oriented perpendicular to the *x* axis (i.e. in our simulations: parallel to the GB). $\boldsymbol{I}_i$ is the line vector of a dislocation segment *i* located within the volume $\Delta V(x)$, and $\boldsymbol{b}_i$ is the Burgers vector of the segment. The symbol "< >" denotes an average over multiple simulations. According to Zhang et al. [44], only two components of the GND tensor, $\alpha_{23}$ and $\alpha_{32}$, are related to the normal plastic strain gradient along the *x* axis which we characterize here in terms of the values of $\alpha_{23}$ and $\alpha_{32}$, averaged over thin slices across the sample taken parallen to the GB. Fig. 11 shows contour plots of this quantity versus the position *x* and indentation depth *h* for various GB-indenter distances.

Let us first discuss, for reference, the behavior expected for indentation of a smaple without GB. In that case, the $\alpha_{23}$ and $\alpha_{32}$ components of the GND tensor are expected to be zero directly underneath the indenter tip: Because of symmetry, this is a location of maximum plastic strain, where the strain gradient perpendicular to the indentation direction and accordingly the corresponding GND tensor components must be zero. As we pass underneath the indenter, the strain gradient changes sign and at some distance from the indenter we expect the GND density to reach a maximum, the position of which indicates the characteristic extension of the plastic zone. The GND surrounding the indenter tip create a back stress, which decreases with increasing extension of the plastic zone and may be fully or partly responsible for the indentation size effect.

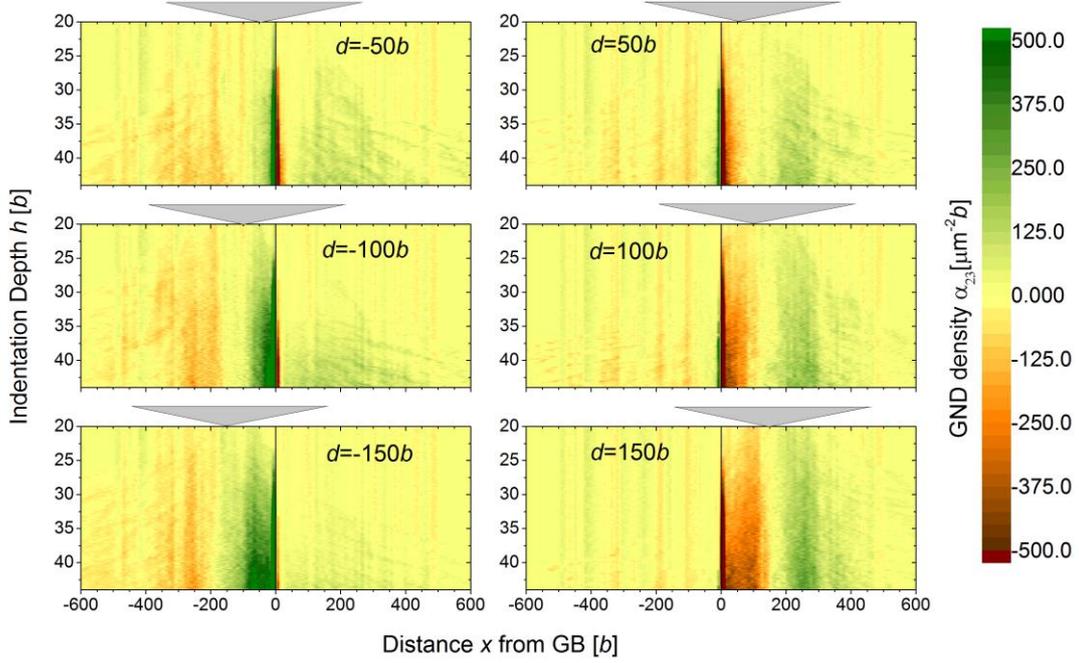

**Fig. 11.** Plots of geometrically necessary dislocation (GND) density tensor component $\alpha_{23}$ versus distance $x$ from GB, for varying indentation depths; the plots represent different distances $d$ from indenter to GB; the position of the indenter is additionally marked with a black triangle.

We now look at the GND distribution observed during indentation close to a GB (Fig. 11). We still find a GND density maximum on the side of the indenter facing away from the GB, which is little affected by the GB-indenter distance. On the side facing the GB, on the other hand, the motion of dislocations is initially constrained to the region enclosed between indenter and GB, and accordingly the GNDs are shifted closer to the indenter and their density in the region enclosed between indenter and GB increases. The corresponding back stresses delimit dislocation activation underneath the indenter and thereby increase the flow stress required to maintain deformation activity within the plastic zone. At the position of the GB ($x=0$), the GND density in all cases reaches a pronounced maximum due to dislocation pile-up. The height of this maximum is delimited by the effect of dislocation transmission, which occurs once the GND piling up against the GB create a critical forward stress sufficient to induce dislocation transmission. The overall scenario of dislocation pile up, dislocation back stresses controlling plastic flow in the active zone, and dislocation transmission, is very similar to the scenario envisaged in the classical pile-up theory of Hall-Petch hardening. We now proceed to formulate this analogy in quantitative terms.

*5.1. Theoretical interpretation: Modified Hall-Petch effect*

To capture the observed GB effect on hardness and the associated GND distribution in quantitative terms, we first need to point out one essential difference between the present situation and the situation in a polycrystal. In polycrystals, a single characteristic length (the grain size) controls the slip distance of dislocations. In our simulations, as in real indentation experiments near GB, this role is shared by the distance from the indenter to the GB on the one hand, and from the indenter to the boundary of the plastic zone on the other hand.

We account for this dual dependency by introducing an effective plastic zone radius via

$$\pi R_{\text{eff}}^2(h,d) = \frac{A(h)}{2}\left[\frac{\eta^2 d^2}{\eta^2 d^2 + A(h)} + 1\right] , \qquad (13)$$

where $\eta$ is a fit parameter and $A(h) \approx 24.5 h^2$ is the projected area underneath the indenter, which we identify with the cross section of the plastic zone in absence of a GB. Eq. (13) is motivated by the following considerations: (i) in the limit $d \to \infty$, the GB effect becomes asymptotically irrelevant and $R_{\text{eff}}$ is equal to the equivalent radius of the area $A$; (ii) in the limit $d=0$, the GB exactly bisects the plastic zone. Hence, the characteristic area over which dislocation loops can expand is reduced exactly by a factor 0.5, from $A$ to $A/2$; (iii) in the general case, Eq. (13) provides a simple analytic interpolation between these limits where the empirical factor $\eta$ determines the relative influence of $d$ on the plastic zone size.

Taking $R_\text{eff}$ to be the effective radius of a circular slip zone in which disloations are piling up tangential to the GB, we can use standard relations of the theory of dislocation pile-ups to derive a modified Hall-Petch relationship for the indentation hardness,

$$H = H_0 + \frac{K_\text{HP}}{\sqrt{R_\text{eff}^*}} \quad . \tag{14}$$

He $R_\text{eff}^* = \sqrt{3} R_\text{eff}$ is the projection of the effective slip zone radius, Eq. (13), which is envisaged parallel to the indented [001] surface, onto a slip plane with normal vector $\mathbf{n} = (1/\sqrt{3})[111]$. The Hall-Petch coefficient $K_\text{HP}$ can then be derived within the framework of pile-up theory. For the parameters in our simulations we obtain $K_\text{HP} = 16.19 \, \text{GPa}\sqrt{b}$, see the derivation of Eq. (A5) in Appendix A. This leaves us with two fit parameters, namely the ratio $\eta$ of slip zone area to indented area, and the asymptotic hardness $H_0$. An excellent fit to the data for all $d$ and $h$ values considered in our simulations is obtained by using the parameters $H_0 = 4.43$ GPa, and $\eta=1.65$. Note that with these parameters, Eqs (13) and (14) account not only for the dependency of hardness on distance between indenter and GB, but also for the dependency on indentation depth. This indicates that, in the regime where indentation is controlled by dislocation nucleation underneath the indenter as envisaged in our simulations, pile-up theories may provide an adequate representation of the associated size effects even if two characteristic length scales (here: $d$ and $h$) are involved.

While our model provides a good representation of the hardness dependence on distance to GB and indentation depth, one may ask whether the assumptions made in deriving the Hall-Petch constant (circular slip zones, stress concentration due to a classical pile-up) are not too strong. It is obvious by simply looking at the simulations that neither the slip zones are exactly circular, nor do the dislocations exactly follow the arrangement expected for a classical pile-up. On the side facing away from the indenter, it would in fact be quite impossible for them to do so, given the absence of a hard boundary.

To check whether our model, despite these caveats, possesses predictive power regarding the distribution of dislocations, we focus on the area of interest between GB and indenter. We characterize the GND density in this area in terms of the averaged GND tensor component magnitude $\rho_{\text{GND}}(V_d) = (1/2)\langle|\alpha_{32}|+|\alpha_{23}|\rangle_{V_d}$, evaluated from Eq. (10) over the volume $V_d = l_y \times l_z \times d$ encompassed between GB ($x=0$) and indenter ($x=d$). The corresponding values of $\rho_{\text{GND}}(V_d)$ were evaluated for all $d$ values studied in our simulations (red circles in Fig. 10) .

The pile-up model can now be used to evaluate the same quantity. The derivation of the GND tensor resulting from a pile-up on a single slip zone is given in the Appendix, with the result given by Eqs. (A13) and (A14). The overall GND density tensor derives from this result as

$$\langle \boldsymbol{\alpha} \rangle_{V_d} = \sum_s \langle \boldsymbol{\alpha}^s \rangle_{V_d} = \rho_\|^s(V_d) \left[ (\mathbf{n}^s \times \mathbf{e}_x) \otimes \mathbf{b}^s \right] \tag{15}$$

where $\rho_\|^s(V_d)$ is the GND line density in slip system $s$ in the direction parallel to the GB, which for the pile-up model used here is for a given slip system calculated according to Eq. (A13). $\mathbf{n}^s$ and $\mathbf{b}^s$ are the respective slip plane normal and Burgers vectors, and the summation runs over all active slip systems.

From Eq. (15) the GND density magnitude $\rho_{\text{GND}}(V_d) = (1/2)\langle|\alpha_{32}|+|\alpha_{23}|\rangle_{V_d}$ corresponding to the pile-up model can be evaluated (red lines in Fig. 10). We find excellent agreement with the simulation data. In particular, the increase of GND density with increasing indentation depth is accurately predicted by the model. In view of the fact that the comparison between model and simulation does not involve any additional fit parameters, it is fair to say the model despite its strong idealization fully captures the interplay between GND density accumulation and indentation size effect. It provides accurate predictions of the influences of both GB-indenter distance and of indentation depth on hardness and dislocation microstructure, and of the accumulation of GNDs in front of the GB. These predictions pertain to the regime of very shallow indents where, as implicit in the initial conditions used in our simulations, plastic activity is controlled by nucleation of dislocation loops underneath the indenter but where bulk dislocation activation is irrelevant.

## 6. Conclusions

In the present work, 3D multiscale DDD has been coupled with a penetrable GB model to simulate nanoindentation of aluminum bicrystals. Initial pop-in and GB pop-in events were observed on the load-depth curves and the relationship between hardness and distance to GB was investigated and interpreted. The main findings are summarized as follows:

(1) A dimensional multiscale modeling framework has been established which couples three-dimensional discrete dislocation dynamics with the finite element method. By considering the physical interaction mechanisms between dislocations and penetrable GBs this famework provides an efficient tool to investigate GB effects on the nanoindentation of bicrystals and polycrystals.

(2) The first pop-in phenomenon in the nanoindentation load-depth occurs at the moment of the first activation or nucleation of dislocations beneath the indenter, indicating the transition from elastic to elasto-plastic response of the indented material. The pop-in events occurring after the initial pop-in are mainly due to dislocation transmission through GB and the intermittent operation of dislocation sources.

(3) Due to the GB hardening effect, the indentation hardness of a bicrystal increases with decreasing distance between indenter and GB. This GB effect superimposes on the indentation size effect related to the plastic zone width and thus the indentation depth. The underlying dislocation microstructure shows that the GND density between indenter and GB increases with decreasing GB-indenter distance, leading to enhancement of GB hardening.

(4) The interplay between GND accumulation and back stresses on the plastic zone can be captured by a pile-up model where the effective radius of a slip zone depends both on indentation depth and on indenter-GB distance. The model leads to a modified Hall-Petch relation which correctly predicts the dependence of hardness on both parameters, and which at the same time allows to quantitatively predict the GND accumulation between GB and indenter. This is particularly remarkable because we observe in this region that increasing indentation depth and decreasing overall hardness go along with *increasing* GND density, a finding which is difficult to reconcile with Taylor-based theories of indentation size effects but which, in our analysis, can be readily explained in the context of a pile-up theory.

(5) The developed model pertains to very shallow nanoindentation, where the volume probed by the indenter is devoid of pre-existing dislocations. In such circumstances, plasticity is controlled by dislocation generation near or at the indenter-sample interface and no bulk dislocation sources or pre-existing bulk dislocations are activated. For experiments carried out under such circumstances, the equations we derive based upon a modified Hall-Petch model may provide a useful tool for analyzing the dependency of hardness on indentation depth and indenter-GB distance.

**Acknowledgements**

This work was supported by National Natural Science Foundation of China (Grant No. 11672251, U1730106); the key R&D program of China (Grant No. 2016YFB1102600) and the Opening fund of State Key Laboratory for Strength and Vibration of Mechanical Structures (Grant No. SV2018-KF-10). M.Zaiser also acknowledges support by the Chinese State Administration of Foreign Experts Affairs under Grant No. MS2016XNJT044. X. Zhang is grateful for support from the Humboldt Research Foundation and to Dr. Minsheng Huang for fruitful discussions and comments on the manuscript.

**Appendix: Some results from the theory of dislocation pile-ups**

A. *Stress concentration on the slip zone boundary and Hall-Petch coefficient*

The theory of dislocation pile-ups in circular and elliptic slip zones was developed by Eshelby [45], using the close analogies with the problem of shear cracks and of ellipsoidal inclusions. Here we mainly refer to the very useful compilation of results by Li and Chou [46]. Consider the displacement across a circular slip zone $S$ of normal vector $\mathbf{n}$ and radius $R$, which is populated by dislocations of Burgers vector $\mathbf{b}$. Assume that the dislocations inside the slip zone are in equilibrium under a shear stress $\tau$, and that the critical flow stress for a dislocation to move is $\tau_0$. The displacement field across the slip zone then has the radially symmetric form

$$\mathbf{u}(r) = \frac{\mathbf{b}}{b} u(r), \quad u(r) = \frac{8(\tau-\tau_0)(1-\nu)}{\pi\mu(2-\nu)} \sqrt{R^2 - r^2} \tag{A1}$$

where $r$ is the radial coordinate [46]. The corresponding dislocation lines can be identified with the circular contour lines where the displacement field has the values $k\mathbf{b}$ and $k$ is an integer number. The number of dislocations in the pile up is

$$n = \frac{8(1-\nu)(\tau-\tau_0)}{\pi(2-\nu)G} \frac{R}{b} \tag{A2}$$

and the average stress acting on the outermost dislocation (the dislocation located at thhe boundary of the slip zone) is given by

$$\tau_n = \frac{n(\tau-\tau_0)}{2} \tag{A3}$$

To assess the critical shear stress that must act in the slip zone for the outermost dislocation to be transmitted through a GB at the boundary of the slip zone, we equate $\tau_n$ to the critical resolved shear stress $\tau_{\text{CRSS}}$ for grain boundary passing. Using (A2,A3) and resolving with respect to $\tau$, we obtain the Hall-Petch type relationship

$$\tau = \tau_0 + \left[\frac{G\pi(2-\nu)\tau_{\text{CRSS}}}{4(1-\nu)}\right]^{1/2}\sqrt{\frac{b}{R}} \tag{A4}$$

With Eq. (10) of the main paper, a GB misorientation angle of $\theta = 5°$ and the parameters of Table 1 we find $\tau_{\text{CRSS}} \approx G/38 \approx 700\,\text{MPa}$. A Hall-Petch coefficient for indentation is then obtained from Eq. (A4) as follows: First we multiply by an effective Schmid factor which for the near-[100] orientation of the indentation direction used in our simulations we take to be $M = \sqrt{6}$, to obtain an equivalent relation for an axial yield stress $\sigma$. Next, we multiply with a constraint factor $C$ to convert this yield stress into a hardness value $H = C\sigma$.

In applying this procedure a word of caution is appropriate. The constraint factor as introduced by Tabor [47] accounts for the fact that the material within the plastic zone pushes against the surrounding, plastically undeformed material and accordingly experiences a back stress in form of a hydrostatic pressure. This pressure, in turn, reduces the shear stress driving plastic deformation, which therefore can only proceed if an enhanced force is exerted on the indenter. For metals where the bulk axial yield stress is much less than the elastic modulus, a calculation assuming ideally plastic behavior and an isotropic Von Mises type yield surface leads to a constraint factor $C \approx 3$ which relates the hardness to the uniaxial yield stress.

However, in the present case the assumptions of ideally plastic behavior and plastic isotropy are surely unwarranted. In fact, the entire model outlined above is built on the assumption that plasticity is confined to platelet-like slip zones which follow crystallographic slip planes, and in which significant back stresses are building up, indicating kinematic hardening. Furthermore, one might argue that the model already accounts for the back stress of dislocations pushing against the boundary of the plastic zone, and that the inclusion of a constraint factor runs into the danger of counting these stresses twice.

It is therefore fortuitous that an actual evaluation of the constraint factor for the present simulations demonstrate that this factor is irrelevant even if we take it into account. To this end, we evaluate the relationship between hardness and axial yield stress using the results of Johnson [48] who obtains for a conical indenter:

$$\frac{H}{\sigma} = \frac{2}{3}\left[1 + \ln\left(\frac{E}{\sigma}\frac{\tan\theta}{3}\right)\right] . \qquad (A5)$$

For a conical indenter that produces for a given indentation depth the same indentation area as a Berkovich indenter, $\theta \approx 19.7°$. The Young's modulus of Al as assumed in the present simulations is $E = 76$ GPa. In the present simulations, typical asymptotic hardness values amount to $H \approx 4$ GPa which then leads, with Eq. (A5), to the typical values $\sigma = 4.28$ GPa, $C = H/\sigma = 1.17$. We thus find that, under the conditions prevailing in the present simulations, the confinement factor is close to unity and indeed very different from its bulk value. The reason for this discrepancy with naïve expectation is clear: Here we are indenting a material where bulk dislocation activity is absent and plasticity is almost entirely controlled by dislocation nucleation underneath the indenter. Accordingly, the equivalent axial yield stress is no longer very small as compared to the elastic modulus, and deformation in our simulations is mainly elastic and not plastic. The theoretical calculation of the confinement factor confirms what intuition leads to expect: Back stresses are already accounted for by the pile-up formalism, and the additional consideration of plastic confinement is not actually of major impact on the results.

In the following we use the above calculated confinement factor $C = 1.17$ which corresponds to the asymptotic hardness we expect for indentation distant from a GB, and which is close to the lower limit $C \approx 1.1$ below which the calculation of a plastic confinement factor becomes meaningless [48]. We emphasize that our results would change little if this factor were left out altogether.

The Hall-Petch-coefficient for the GB effect on indentation then follows as

$$H = K_0 + \frac{K_{HP}}{\sqrt{R}} \quad , \quad K_{HP} = MC\left[\frac{G\pi(2-\nu)\tau_{CRSS}}{4(1-\nu)}\right]^{1/2}\sqrt{b} = 16.19\sqrt{b}\,\text{GPa} \quad . \quad (A6)$$

## B. GND density in a slip zone segment

To derive the spatial distribution of the GND tensor corresponding to a circular slip zone, we start from Eq. (A1) and note that the corresponding plastic distortion has the form

$$\boldsymbol{\beta}_p(\mathbf{r}) = (\mathbf{n}\otimes\mathbf{u}(\mathbf{r}))\delta_S(\mathbf{r}) \tag{A7}$$

where the integral of the delta-like distribution $\delta_S(\mathbf{r})$ over an arbitrary volume $V$ gives the area of intersection between the slip zone $S$ and the volume $V$.

$$A_{S\cap V} = \int_V \delta_S(\mathbf{r})\,\mathrm{d}^3 r \tag{A8}$$

We may now evaluate the average over the volume $V$ of GND tensor related to the slip distribution (A1) in $S$. To this end, we use that the Burgers vector $\boldsymbol{b}$ and hence the displacement $\boldsymbol{u}$ are parallel to $S$ and introduce a slip zone coordinate system $\tilde{x}, \tilde{y}, \tilde{z}$, where the slip zone is located in the plane $\tilde{z} = 0$ (hence, the unit vector $\mathbf{e}_{\tilde{z}} = \mathbf{n}$), the $\tilde{y}$ axis is aligned with the intersection between the slip plane (normal vector $\mathbf{n}$) and the GB (normal vector $\mathbf{e}_x$), hence $\mathbf{e}_{\tilde{y}} = \mathbf{n}\times\mathbf{e}_x/|\mathbf{n}\times\mathbf{e}_x|$, and the unit vector in $\tilde{x}$ direction is $\mathbf{e}_{\tilde{x}} = \mathbf{n}\times\mathbf{e}_x\times\mathbf{n}/|\mathbf{n}\times\mathbf{e}_x|$. In this coordinate system, the non-vanishing components of the dislocation density tensor are

$$\begin{aligned}\alpha_{\tilde{x}\tilde{y}} &= (b_{\tilde{y}}/b)\partial_{\tilde{y}} u\,\delta_S(\mathbf{r}) \quad , \quad \alpha_{\tilde{y}\tilde{x}} = (b_{\tilde{x}}/b)\partial_{\tilde{x}} u\,\delta_S(\mathbf{r}) \\ \alpha_{\tilde{x}\tilde{x}} &= (b_{\tilde{x}}/b)\partial_{\tilde{y}} u\,\delta_S(\mathbf{r}) \quad , \quad \alpha_{\tilde{y}\tilde{y}} = (b_{\tilde{y}}/b)\partial_{\tilde{x}} u\,\delta_S(\mathbf{r})\end{aligned} \tag{A9}$$

We now consider a cuboidal volume $V_d$ of extension $d*l_y\times l_z$, which intersects the slip plane in such a manner that one of its sides is tangential to the slip zone and the area of intersection $A_{S\cap V}$ has the form of a circle segment as shown in Fig. A1.

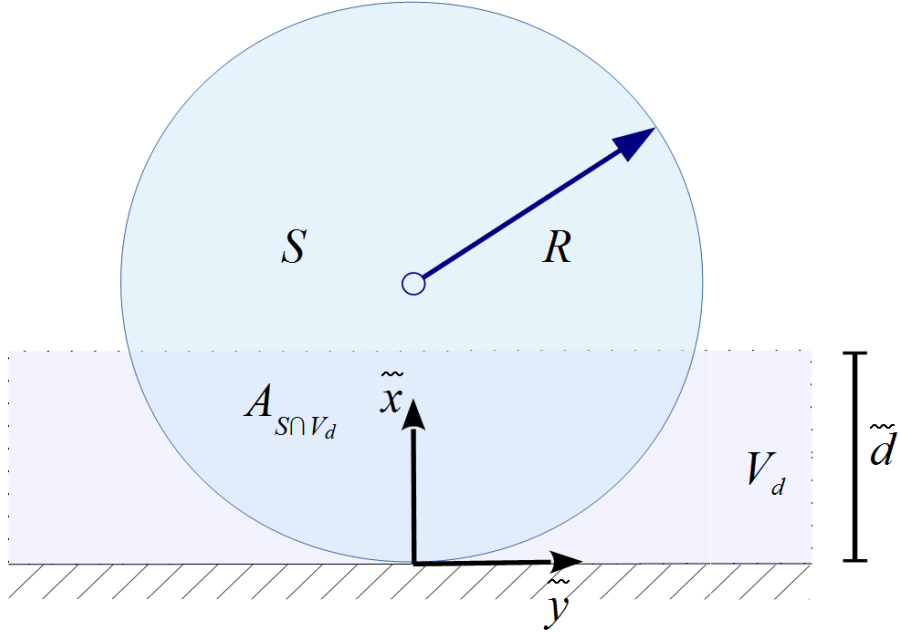

**Fig. A1:** Intersection of a circular slip zone $S$ with the volume $V_d$ contained between grain boundary and indenter. The coordinate axes represent slip system coordinates that are in general inclined with respect to the system coordinates; $\tilde{d}$ is the projection of the GB-indenter distance $d$ onto the slip plane.

We now evaluate the dislocation density tensor averaged over the volume $V_d$ as

$$\langle \alpha_{\tilde{y}\tilde{x}} \rangle_{V_d} = \rho_{\tilde{x}}(V_d) b_{\tilde{x}}, \quad \langle \alpha_{\tilde{y}\tilde{y}} \rangle_{V_d} = \rho_{\tilde{x}}(V_d) b_{\tilde{y}}, \quad . \tag{A10}$$

where

$$\rho_{\tilde{y}}(V_d) = \frac{1}{bdl_y l_z} \int_{A_{S \cap V_d}} \partial_{\tilde{x}} u(r) \, d\tilde{x} d\tilde{y} \tag{A11}$$

can be understood as the average dislocation line density (line length per unit volume) in $\tilde{y}$ direction. Note that the two components $\alpha_{\tilde{x}\tilde{y}}$ and $\alpha_{\tilde{x}\tilde{x}}$, which represent line components aligned with the $\tilde{x}$ direction and involve integration of the $\tilde{y}$ derivative of $u$, vanish upon integration due to symmetry. The integration in (A11) can be performed analytically using Eq. (A1), giving the simple result

$$\rho_{\tilde{y}}(V_d) = \frac{1}{bdl_yl_z}\frac{4(\tau-\tau_0)(1-\nu)}{G(2-\nu)}\left(2R\tilde{d}-\tilde{d}^2\right) \ . \tag{A12}$$

For a general slip zone that is not perpendicular to the boundary of $V_d$, the projected segment width fulfills the relation $\tilde{d}^2 = d^2(1-(\mathbf{n}\cdot\mathbf{e}_x)^2)^{-1/2} = \lambda^2 d^2$ which for $1/\sqrt{3}[111]$ slip planes intersecting the plane ($x=0$) in a fcc structure gives the result $\lambda = \sqrt{3/2}$. We may now combine Eq. (A12) with (A4) and (A5) to express the averaged GND line density as

$$\rho_{\tilde{y}}(V_d) = \frac{1}{l_yl_z}\frac{K_{HP}}{CG}\frac{(1-\nu)}{(2-\nu)}\frac{4R-d\sqrt{6}}{b}\sqrt{\frac{b}{R}} \tag{A13}$$

From this result, the GND tensor due to the considered slip zone derives as

$$\langle\boldsymbol{\alpha}\rangle_{V_d} = \rho_{\tilde{y}}(V_d)\left[\mathbf{e}_{\tilde{y}}\otimes\mathbf{b}\right] = \rho_{\tilde{y}}(V_d)\left[(\mathbf{n}\times\mathbf{e}_x)\otimes\mathbf{b}\right] \ . \tag{A14}$$

## References

[1] J.R. Greer and J.T.M. De Hosson, Progress in Materials Science, Plasticity in Small-Sized Metallic Systems: Intrinsic versus Extrinsic Size Effect. Progr. Mater. Sci. 56 (2011) 654-724.
[2] D. Kiener, R. Pippan, C. Motz, H. Kreuzer, Microstructural evolution of the deformed volume beneath microindents in tungsten and copper, Acta Mater. 54(10) (2006) 2801-2811.
[3] F. Javaid, E. Bruder, K. Durst, Indentation size effect and dislocation structure evolution in (001) oriented SrTiO 3 Berkovich indentations: HR-EBSD and etch-pit analysis, Acta Mater. 139 (2017) 1-10.
[4] W.C. Oliver, G.M. Pharr, Improved technique for determining hardness and elastic modulus using load and displacement sensing indentation experiments, J. Mater. Res. 7(06) (1992) 1564-1583.
[5] H. Fan, Z. Li, M. Huang, Toward a further understanding of intermittent plastic responses in the compressed single/bicrystalline micropillars, Scripta Mater. 66(10) (2012) 813-816.
[6] W.A. Soer, Interactions between dislocations and grain boundaries, Groningen: University of Groningen, 2016.
[7] W.A. Soer, K.E. Aifantis, J.T.M.D. Hosson, Incipient plasticity during nanoindentation at grain boundaries in body-centered cubic metals, Acta Mater. 53(17) (2005) 4665-4676.
[8] G.Z. Voyiadjis, C. Zhang, The mechanical behavior during nanoindentation near the grain boundary in a bicrystal FCC metal, Mat. Sci. Eng. A 621 (2015) 218-228.
[9] C. Zhang, G.Z. Voyiadjis, Rate-dependent size effects and material length scales in nanoindentation near the grain boundary for a bicrystal FCC metal, Mater. Sci. Eng. A 659 (2016) 55-62.
[10] M. Liu, C. Lu, K.A. Tieu, K. Zhou, Crystal plasticity FEM study of nanoindentation behaviors of Cu bicrystals and Cu–Al bicrystals, J. Mater. Res. 30(16) (2015) 2485-2499.

[11] C. Kirchlechner, F. Toth, F.G. Rammerstorfer, F.D. Fischer, G. Dehm, Pre- and post-buckling behavior of bi-crystalline micropillars: Origin and consequences, Acta Mater. 124 (2017) 195-203.

[12] M.C. Fivel, C.F. Robertson, G.R. Canova, L. Boulanger, Three-dimensional modeling of indent-induced plastic zone at a mesoscale 1, Acta Mater. 46(17) (1998) 6183-6194.

[13] Y.L. Chiu, A.H.W. Ngan, A TEM investigation on indentation plastic zones in Ni 3 Al(Cr,B) single crystals, Acta Mater. 50(10) (2002) 2677-2691.

[14] J.L. Zhang, S. Zaefferer, D. Raabe, A study on the geometry of dislocation patterns in the surrounding of nanoindents in a TWIP steel using electron channeling contrast imaging and discrete dislocation dynamics simulations, Mater. Sci. Eng. A 636 (2015) 231-242.

[15] L. Li, L. Shen, G. Proust, C.K.S. Moy, G. Ranzi, Three-dimensional crystal plasticity finite element simulation of nanoindentation on aluminium alloy 2024, Mat. Sci. Eng. A 579 (2013) 41-49.

[16] A. Widjaja, A. Needleman, E. Van der Giessen, The effect of indenter shape on sub-micron indentation according to discrete dislocation plasticity, Model. Simul. Mater. Sc 15(1) (2006) S121-S131(11).

[17] A. Widjaja, E. Van der Giessen, V.S. Deshpande, A. Needleman, Contact area and size effects in discrete dislocation modeling of wedge indentation, J. Mater. Res. 22(3) (2007) 655-663.

[18] A. Needleman and E. Van der Giessen, Discrete dislocation plasticity: a simple planar model. Model. Simul. Mater Sci Eng 3 (1999) 689-735.

[19] A. Widjaja, E. Van der Giessen, A. Needleman, Discrete dislocation analysis of the wedge indentation of polycrystals, Acta Mater. 55(19) (2007) 6408-6415.

[20] C. Ouyang, Z. Li, M. Huang, C. Hou, Discrete dislocation analyses of circular nanoindentation and its size dependence in polycrystals, Acta Mater. 56(12) (2008) 2706-2717.

[21] J. Hu, Z. Liu, Y. Cui, F. Liu, Z. Zhuang, A new view of incipient plastic instability during nanoindentation, Chinese. Phys. Lett. 34 (2017) 64-67.

[22] G. Po, M.S. Mohamed, T. Crosby, C. Erel, A. El-Azab, N. Ghoniem, Recent progress in discrete dislocation dynamics and its applications to micro plasticity, J. Min. Met. Mat. S. 66(10) (2014) 2108-2120.

[23] H.J. Chang, M. Fivel, D. Rodney, M. Verdier, Multiscale modelling of indentation in FCC metals: From atomic to continuum, Cr. Phys. 11(3) (2010) 285-292.

[24] J.A. El-Awady, S.B. Biner, N.M. Ghoniem, A self-consistent boundary element, parametric dislocation dynamics formulation of plastic flow in finite volumes, J. Mech. Phys. Solids. 56(5) (2008) 2019-2035.

[25] H.M. Zbib, M. Rhee, J.P. Hirth, On plastic deformation and the dynamics of 3D dislocations, Int. J. Mech. Sci. 40(2) (1998) 113-127.

[26] H.M. Zbib, T. Diaz de la Rubia, A multiscale model of plasticity, Int. J. Plasticity. 18(9) (2002) 1133-1163.

[27] O. Jamond, R. Gatti, A. Roos, B. Devincre, Consistent formulation for the discrete-continuous model: improving complex dislocation dynamics simulations, Int. J. Plasticity 80 (2016) 19-37.

[28] M. Huang, S. Liang, Z. Li, An extended 3D discrete-continuous model and its application on single- and bi-crystal micropillars, Model. Simul. Mater. Sc 25 (2017).

[29] C. Lemarchand, B. Devincre, L.P. Kubin, Homogenization method for a discrete-continuum simulation of dislocation dynamics, J. Mech. Phys. Solids. 49(9) (2001) 1969-1982.

[30] Z.L. Liu, X.M. Liu, Z. Zhuang, X.C. You, A multi-scale computational model of crystal plasticity at submicron-to-nanometer scales, Int. J. Plasticity. 25(8) (2009) 1436-1455.

[31] A. Vattré, B. Devincre, F. Feyel, R. Gatti, S. Groh, O. Jamond, A. Roos, Modelling crystal plasticity by 3D dislocation dynamics and the finite element method: The Discrete-Continuous Model revisited, J. Mech. Phys. Solids. 63(2) (2014) 491-505.

[32] W. Cai, A. Arsenlis, C.R. Weinberger, V.V. Bulatov, A non-singular continuum theory of dislocations, J. Mech. Phys. Solids. 54(3) (2006) 561-587.

[33] G.C. Hasson, C. Goux, Interfacial energies of tilt boundaries in aluminium. Experimental and theoretical determination, Scripta Metallurgica 5(10) (1971) 889-894.

[34] Z. Li, C. Hou, M. Huang, C. Ouyang, Strengthening mechanism in micro-polycrystals with penetrable grain boundaries by discrete dislocation dynamics simulation and Hall–Petch effect, Comp. Mater. Sci. 46(4) (2009) 1124-1134.

[35] L.E. Murr, Some observations of grain boundary ledges and ledges as dislocation sources in metals and alloys, Metallurgical Transactions A 6(3) (1975) 505-513.

[36] G.N. Peggs, I.C. Leigh, Recommended procedure for micro-indentation vickers hardness test, Report MOM 62, UK National Physical Laboratory, 1983.

[37] M. Yaghoobi and G.Z. Voyiadjis, Large-scale atomistic simulation of size effects during nanoindentation: Dislocation length and hardniess, Materials Science and Engineering A 634 (2015) 20-31.

[38] X. Zhang, K.E. Aifantis, Interpreting strain bursts and size effects in micropillars using gradient plasticity, Mater. Sci. Eng. A 528(15) (2011) 5036-5043.

[39] X. Zhang, K.E. Aifantis, Accounting for grain boundary thickness in the sub-micron and nano scales, Rev. Adv. Mater. Sci. 26(1) (2010) 74-90.

[40] D. Lorenz, A. Zeckzer, U. Hilpert, P. Grau, H. Johansen, H.S. Leipner, Pop-in effect as homogeneous nucleation of dislocations during nanoindentation, Phys. Rev. B 67(17) (2003) 386-393.

[41] H. Bei, Y.F. Gao, S. Shim, E.P. George, G.M. Pharr, Strength differences arising from homogeneous versus heterogeneous dislocation nucleation, Phys. Rev. B 77(6) (2008) 439-446.

[42] M.G.Wang, A.H.W. Ngan, Indentation strain burst phenomenon induced by grain boundaries in niobium, J. Mater. Res. 19(8) (2004) 2478-2486.

[43] K.E. Aifantis, J. Senger, D. Weygand, M. Zaiser, Discrete dislocation dynamics simulation and continuum modeling of plastic boundary layers in tricrystal micropillars, IOP Conference Series: Mater. Sci. Eng. 3(1) (2009):012025.

[44] X. Zhang, K.E. Aifantis, J. Senger, D. Weygand, M. Zaiser, Internal length scale and grain boundary yield strength in gradient models of polycrystal plasticity: How do they relate to the dislocation microstructure? J. Mater. Res. 29(18) (2014) 2116-2128.

[45] J.D. Eshelby, The distribution of dislocations in an elliptical glide zone, physica status solidi 3 (1963), 2057-2060.

[46] J.C.M. Li and Y.T. Chou, The role of dislocations in the flow stress grain size relationships, Metall. Trans 1 (1970), 1145-1159

[47] D. Tabor, Hardness of Metals, Clarendon Press, Oxford 1951.

[48] K. L. Johnson, The correlation of indentation experiments, J. Mech. Phys. Solids 18 (1970) 115-126.